\documentclass[12pt]{iopart}

\usepackage{graphicx}
\usepackage{amssymb}
\usepackage{epstopdf}
\usepackage{subfig}

\begin{document}

\title{Phenotypic spandrel: absolute discrimination and ligand antagonism}

\author{Paul Fran\c cois}%
\author{Mathieu Hemery}
\author{Kyle A. Johnson}

\author{Laura N. Saunders}

\address{Physics Department, McGill University, Montreal, Quebec, Canada H3A 2T8}%


\begin{abstract}

We consider the general problem of sensitive and specific discrimination between biochemical species.  An important instance is immune discrimination between self and not-self, where it is also observed experimentally that ligands just below discrimination threshold negatively impact response, a phenomenon called antagonism. We characterize mathematically the generic properties of such discrimination, first relating it to biochemical adaptation. Then, based on basic biochemical rules, we establish that, surprisingly, antagonism is a generic consequence of {\it any} strictly specific discrimination made independently from ligand concentration. Thus antagonism constitutes a ``phenotypic spandrel": a phenotype existing as a necessary by-product of another phenotype.  We exhibit a simple analytic model of discrimination displaying antagonism, where antagonism strength is linear in  distance from detection threshold.  This contrasts with traditional proofreading based models where antagonism vanishes far from threshold and thus displays  an inverted hierarchy of antagonism compared to simpler models. The phenotypic spandrel studied here is expected to structure many decision pathways such as immune detection mediated by TCRs and  FC$\epsilon$RIs, as well as endocrine signalling/disruption.
\end{abstract}

\maketitle

\section*{Introduction}

 Recent works in quantitative evolution combined to mathematical modelling have shown that evolution of biological networks is constrained by selected phenotypes in strong unexpected ways.  Trade-offs between different functionalities increasingly appear as major forces shaping evolution of complex phenotypes moving on evolutionary Pareto fronts \cite{Shoval:2012ke, Warmflash:2012}. Numerical experiments of {\it in silico} evolution of phenotypic models of gene networks  \cite{Francois:2014} have further shown that, surprisingly, selection of complex phenotypes leads to apparition of complex traits that have not been explicitly selected for. Such phenomena are reminiscent of the architectural  image of ``evolutionary spandrels" proposed by Gould and Lewontin  \cite{Gould:1979vo}. They argued that many biological properties are necessary by-products of more fundamental adaptive traits, due to underlying constraints (e.g. tridimensional geometry  in the case of architectural spandrels). Spandrels can themselves be tinkered by evolution into new functional structures, leading to the notion of ``exaptation"\cite{Gould:1982ta}. But we are still lacking a quantitative theory of such spandrels, which might explain that other scholar have questioned the notion of spandrel (see a summary in Gould's own rebuttal \cite{Gould:1997ue}).

A biological example of broad interest is the absolute discrimination between different molecule ``qualities'', an instance being early immune recognition of antigens by T cells \cite{Malissen:1998dy,Feinerman:2008b,Lever:2014, Chakraborty:2014hw}. Distinction performed by T cells between self and not-self is based on a measure of some effective biochemical parameter $\mu$ characteristic of ligand quality. The quality of an antigen has been first suggested to be defined by the typical binding time  $\tau$ \cite{Kersh:1998a,Gascoigne:2001} of ligands to T cell receptors (TCRs), defining the so-called ``life-time dogma" \cite{Feinerman:2008b}. For larger association rate, the affinity $K_D$ has also been proposed to be the parameter discriminating between self and not-self  \cite{Dushek:2009fi,Govern:2010kx,Aleksic:2010hh}  . Discrimination has to be extremely specific to quality $\mu$ (e.g. to prevent an auto-immune disease), insensitive to the very high number of self ligands  with $\mu<\mu_c$,  but also very sensitive to very low ligand concentrations (e.g. to detect a nascent infection), thus the term ``absolute discrimination" \cite{Francois:2016ig}.  Recent works in immunology have confirmed that discrimination by T cells is almost absolute, and furthermore shown how it can be modulated by external cytokines such as IL-2 \cite{Voisinne:2015}.

In evolutionary simulations aiming at reverse-engineering absolute discrimination \cite{Lalanne:2013}, evolved networks always present undesirable  ligand antagonism. Antagonism is a ``dog in the manger" effect, due to the cross-talk between different ligands: like the dog in Aesop's fable who can not eat the grain but nevertheless prevents the horse to eat anything either, ligands that are unable to trigger response actively prevent agonists to do so. Such effects have been described experimentally in several immune recognition processes \cite{Torigoe:1998vj,AltanBonnet:2005}. It is intriguing that antagonism similar to nature spontaneously appears in evolutionary simulations without explicit selection. In both simulated and real networks, antagonism is a consequence of crucial negative interactions for specific and sensitive ligand detection \cite{Dittel:1999,AltanBonnet:2005,Francois:2013,Lalanne:2013}. We could show that the main effect of kinetic proofreading in this system is to lower antagonism \cite{Lalanne:2013}, but is it possible to construct an absolute discrimination system without any antagonism ? In the following, we show from general arguments how  ligand antagonism is a necessary consequence of absolute discrimination, thus qualifying as a ``phenotypic spandrel". Our approach allows us to characterize generic solutions of absolute discrimination, to exhibit minimum networks and families of absolute discrimination models and to  further show how addition of proofreading steps leads to an ``inverted" hierarchy of antagonism that is {\it not} a generic feature of the antagonism spandrel. Finally we generalize this theory by making connections with other biochemical examples, including endocrine signalling.

\section*{Antagonism is a consequence of absolute discrimination}

\begin{figure}[h]
\includegraphics[width=\textwidth]{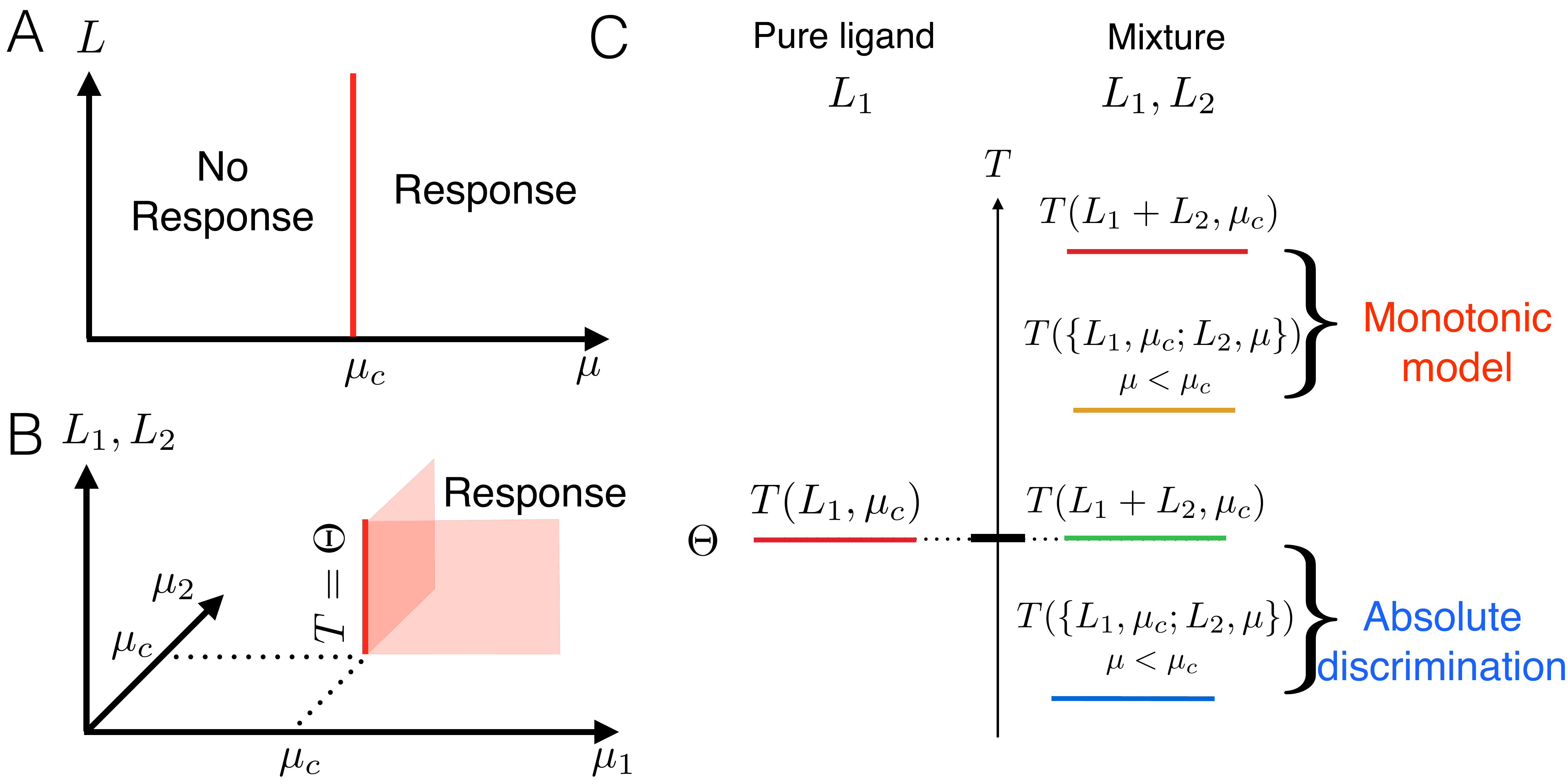}
 \caption{(A). Sketch of absolute discrimination for a single ligand type in $(L,\mu)$ plane. Discrimination line corresponding to $\mu=\mu_c$ is in red. (B). Sketch of absolute discrimination for mixtures, adding a second type of ligands with quality $\mu_2$, with $L_1,L_2$ ligand concentrations represented by a vertical axis. The discrimination line for $\mu_1=\mu_2=\mu_c$ is vertical and satisfies equation \ref{Adaptation}. Red planes correspond to mixtures of agonists where necessary $T> \Theta$.  (C). Intuitive explanation of antagonism (see main text). Different color lines correspond to different output $T$ levels for different Inputs. On the left we represent the reference value $\Theta$ of the output for different models when $L_1$ ligands at $\mu=\mu_c$ are presented. On the right, we look at the variation of the Output when we present a mixture of the same $L_1$ ligands with $L_2$ ligands with different qualities. Monotonic model corresponds to warm colours (red  for both ligands of same quality $\mu=\mu_c$ and orange for different ligand qualities), and always yield an Input increase irrespective of the quality of the added $L_2$ ligands (assuming decision is made for higher Output values). Absolute discrimination  corresponds to cold colours (green for both ligands of same quality $\mu=\mu_c$  and blue  for different ligand qualities). For absolute discrimination, any amount of ligand at $\mu_c$ imposes that the Output remains on the discrimination line ($T=\Theta$, dotted line), but if $\mu<\mu_c$, relative contribution of $L_2$ ligands is negative due to the monotonicity in $\mu$.}  \label{fig:f1}
\end{figure}

 We consider a family of biochemical ligands, with different quantitative properties encoded by a continuous parameter $\mu$, subsequently called ``ligand quality''. Ideal absolute discrimination is performed when cells successfully discriminate between  two categories of ligands, above and below  critical quality  $\mu_c$, irrespective of ligand quantity  $L$. An idealized response diagram for absolute discrimination in $(L,\mu)$ plane defines a vertical  ``discrimination line" at $\mu_c$ (Fig \ref{fig:f1} A) . We model absolute discrimination using  biochemical networks with  continuous, single-valued variables, at steady state, in a way similar to phenotypic models such as the ones reviewed in \cite{Lever:2014}. Ligands interact with receptors at the surface of the cell, we make a mean-field approximation so that all concentrations are averaged out over one cell, and all variables are assumed to be continuous and differentiable (due to biochemistry).
  
We assume that a cellular network is responsible for discrimination between ligand qualities, so that decision is eventually based on thresholding some internal variable $T$ (that could for instance be the steady state value of a downstream protein concentration, or some total phosphorylation state). We however require $T$ to be a differentiable function of parameters because of biochemistry. We will call $T(L,\mu)$ the value of $T$ when $L$ ligands of single quality $\mu$ are presented. While $L$ is in principle a discrete variable, it can be treated as a continuous one for our calculations without loss of generality.

  {\it In vivo}, cells are exposed to complex mixtures of ligands interacting with receptors at the surface of a single cell.  We will use notation $T(\{L_n,\mu_n\})$ for the output value in presence of a mixture, where $L_n$ is the number of ligands presented of quality $\mu_n$ , and again insist on differentiability of this variable. Quality of ligands are thus indexed by $n$, e.g. $n=1,2$ if only two types of ligands are presented, $L_1$ ligands of quality $\mu_1$ and $L_2$ ligands of quality $\mu_2$. We will mostly limit ourselves to mixture of two ligands but our reasoning can be easily generalized. 

We start by connecting the mixture case $T(\{L_n,\mu_n\})$ to the pure case, assuming $\mu_n=\mu_c+d\mu_n$ with very small $d\mu_n$.To build some intuition about the result,  let us start with an ensemble of $L$ identical ligands at $\mu_c$ and vary  the quality of {\bf one single ligand} presented to the cell by an infinitesimal quantity $d\mu_1$. This should translate into an infinitesimal change of the output variable $T(L,\mu_c)$ at linear order due to change of quality $d\mu_1$ of a single ligand, defining the quantity $\mathcal{A}$:

\begin{equation}
  T(\{L-1,\mu_c;1,\mu_c+d\mu_1\})=T(L,\mu_c)+\mathcal{A} d\mu_1
\end{equation}
$\mathcal{A}$ is a function of $L$ and $\mu_c$, and is well defined in a mean-field approximation where all ligands are equivalent and  $T$  is differentiable as required.

Now let us consider another individual ligand. If we vary its quality by an infinitesimal quantity $d\mu_2$,  of same order of magnitude of $d\mu_1$, at linear order this adds another contribution to $T$ which is equal to $\mathcal{A}(L,\mu_c) d\mu_2$. This is the same $\mathcal{A}(L,\mu_c)$ because all ligands are equivalent, and even though the first ligand changed of quality by $d\mu_1$, this only infinitesimaly changes all variables in the system by terms of order $d\mu_1$ so that at lowest order there is no change for $\mathcal{A}$.



We can then generalize this reasoning to any number of ligands: each infinitesimal variation of quality $d\mu_n$ of one single ligand gives an equivalent infinitesimal contribution $\mathcal{A}(L,\mu_c) d\mu_n$ at linear order, so that considering a mixture $\{L_n,\mu_n\}$ we get directly (factorizing $\mathcal{A}$)

  \begin{eqnarray}
    T(\{L_n,\mu_n=\mu_c+d\mu_n \})&=&  T(L,\mu_c) +\mathcal{A}(L,\mu_c) \sum_n L_n d\mu_n\label{Taylor}
  \end{eqnarray}
calling $L=\sum_n L_n$. This is a completely generic result for mixtures that does not depend on the fact that the system is doing absolute discrimination.

As a simple example, let us assume a simple ligand-receptor system far from saturation, where ligands bind to receptors with binding time $\mu^{-1}$,  with a local kinetic amplification mechanism (e.g. proofreading \cite{Mckeithan:1995}), so that $T_n= \kappa L_n \mu_n^2$ is the average number of receptors bound to the ligands of quality $\mu_n$, assuming $L_n$ ligands of this quality are presented, and $\kappa$ some reaction rate (for simplicity we take $\kappa=1$ in the following). Let us consider as an output the total number of receptors presented $T=\sum_n T_n$. Then following our reasoning, starting with $L$ ligands at $\mu_c$ so that $T= L \mu_c$, changing the quality of one ligand by $d\mu_1$ gives a total number of bound receptors $T=(L-1)\mu_c^2 + 1(\mu_c+d\mu_1)^2 \simeq L \mu_c^2 + 2 \mu_c d\mu_1$ at linear order, i.e. with our notation $\mathcal{A}=2 \mu_c$. Changing the quality of another ligand by $d\mu_2$, we get $T=(L-2)\mu_c^2 + 1(\mu_c+d\mu_1)^2+ 1(\mu_c+d\mu_2)^2 \simeq L \mu_c^2+2 \mu_c (d\mu_1+d\mu_2)$. This reasoning can clearly be generalized to get equation \ref{Taylor}. While this model looks extremely simple, the more complex examples with feedback presented in the following work in a similar way.

Coming back to our general reasoning, we focus now on  absolute discrimination, where detection of ligand quality $\mu$ is specific and done independently from total ligand concentration $L$. We assume some response is triggered if  $T\geq\Theta$.
For $L$ ligands at $\mu=\mu_c$ from Figure \ref{fig:f1}A, by continuity we necessarily have on the discrimination line:

\begin{equation}
 \forall L \quad T(L,\mu_c)=\Theta \label{Adaptation}
\end{equation}

So $T$ is independent from the concentration of ligand $L$ when $\mu=\mu_c$, or in other word is a biochemically adaptive variable as a function of $L$ \cite{Francois:2008,Ma:2009}. Further assuming decision is made if $\mu>\mu_c$, and as a consequence of $T$ getting higher than threshold, we necessarily have $T(L,\mu_c+d\mu)>\Theta=T(L,\mu_c)$.  Using equation \ref{Taylor} with a single type of ligand, we thus get $\mathcal{A}(L,\mu_c)>0$.

Consider now the problem of mixtures of ligands of two kinds, sketched on Fig \ref{fig:f1}B . From equations \ref{Taylor} and positivity of $\mathcal{A}$, if all $d \mu_n$ are positive, then $T(\{L_n,\mu_c+d\mu_n\})>T(L,\mu_c)=\Theta$, which in plain words means that a mixture of agonists (close to threshold) always triggers response. A more interesting case is to consider what happens with two different types of ligands, $L_1$ at $\mu_c$ and $L_2$ at $\mu_c-d\mu$. From \ref{Taylor}, we get immediately

\begin{equation}
T(\{L_1,\mu_c;L_2,\mu_c-d\mu\})=T(L_1+L_2,\mu_c)- L_2 d\mu \mathcal{A}(L_1+L_2,\mu_c) \label{Taylor_anta}
\end{equation}

Now from equation \ref{Adaptation}, we have $T(L_1+L_2,\mu_c)=\Theta=T(L_1,\mu_c)$, and since $\mathcal{A}>0$ we thus have  $T(\{L_1,\mu_c;L_2,\mu_c-d\mu\})-T(L_1,\mu_c)=- L_2 d\mu \mathcal{A}(L_1+L_2,\mu_c) <0$, so that we get our main result:

\begin{equation}
T(\{L_1,\mu_c;L_2,\mu_c-d\mu\})<T(L_1,\mu_c)
\end{equation}

This expression establishes ligand  antagonism: a mixture of sub-threshold ligands $L_2$ with critical agonist ligands $L_1$ yields lower signalling variable $T(\{L_1,\mu_c;L_2,\mu_c-d\mu\})$ compared to the case where the same quantity of ligand $L_1$ only is presented $T(L_1,\mu_c)$. Thus if decision is based on thresholding of output $T$, response disappears. So antagonism is established as a general property of systems performing absolute discrimination based on one parameter $\mu$, completely irrespective of internal biochemistry.

An intuitive explanation of the above reasoning can be made, by comparison with general models for signalling that do {\it not} perform absolute discrimination. Let us consider the behaviour of an output variable $T$ of a given signalling pathway, before any thresholding-based decision. We study how its position varies with respect to a reference level, where only one type of ligands is presented but where $T$ level is so that the cell responds to the external signal. For many models of signalling pathways with independent receptors, we expect that output function $T$ is monotonic in both $L$ and $\mu$. For instance, in kinetic proofreading proposed in \cite{Mckeithan:1995}, each receptor contributes additively to signalling once it is bound, so that more ligands, or with longer binding time,  necessarily gives stronger signal. In such models, starting from a critical ligand concentration $L_1$ triggering response, any addition of ligands $L_2$ thus gives {\it higher} output $T$ value, and thus if response is based on thresholding of $T$, response is maintained. Furthermore, addition of $L_2$ ligands with critical quality $\mu_c$  gives higher output than addition of same quantity $L_2$ of ligands with lower quality $\mu<\mu_c$  (Figure \ref{fig:f1}C).  Considering now absolute discrimination and again addition of $L_2$ ligands on Figure \ref{fig:f1}C, the constraint encoded by adaptation from equation \ref{Adaptation}  means that  addition of extra $L_2$ ligands of quality $\mu=\mu_c$ does not change the value of output $T$. But if discrimination is based on $\mu$, we expect that output $T$ still is a monotonic function of $\mu$. Thus, from this point with $L_1+L_2$ ligands at $\mu_c$,  if ligand quality of the extra $L_2$ ligands is lowered ($\mu<\mu_c$), we still expect a lower contribution of those ligands to the Output $T$, just like the monotonic example (Figure \ref{fig:f1}C). Thus if we started right at threshold for $T$, response is now below threshold of activation, corresponding to antagonism.

Antagonism thus appears as a direct consequence of biochemical adaptation at $\mu=\mu_c$. As long as such adaptive behaviour is observed and variable $T$ is differentiable, antagonism ensues, irrespective of details of biochemistry (such as receptor complexations, non linearities in networks, etc...). Counterexamples can be built if the differentiation hypothesis does not hold: for instance if the cell could measure the maximum of binding times of individual ligands, which is clearly not a differentiable function, then there would be no antagonism. Equation \ref{Taylor} further tells us that antagonism occurs in ligand mixtures at linear order as soon as $\sum L_n d \mu_n<0$. Of course it will also be observed for some range of $d\mu$ even at non-linear order. Furthermore, adaptation does not necessarily have to be perfect: non-monotonic response curves  \cite{Vandenberg:2012fc}, with $T$ varying around $\Theta$, essentially approximate well adaptation necessary for absolute discrimination and display antagonism \cite{Francois:2013, Lalanne:2013}. Mathematically, position with respect to threshold depends on the competition between the``flatness" of $T(L,\mu_c)$ as a function of $L$ and the mixture term $\mathcal{A}(L,\mu_c)\sum_n L_n d\mu_n$. Examples of this effect are given in the next section.

We now exhibit and study two interesting classes of models performing absolute discrimination but with different antagonistic behaviour.

\section*{Simple homeostatic model}

\begin{figure}[h]
\includegraphics[width=\textwidth]{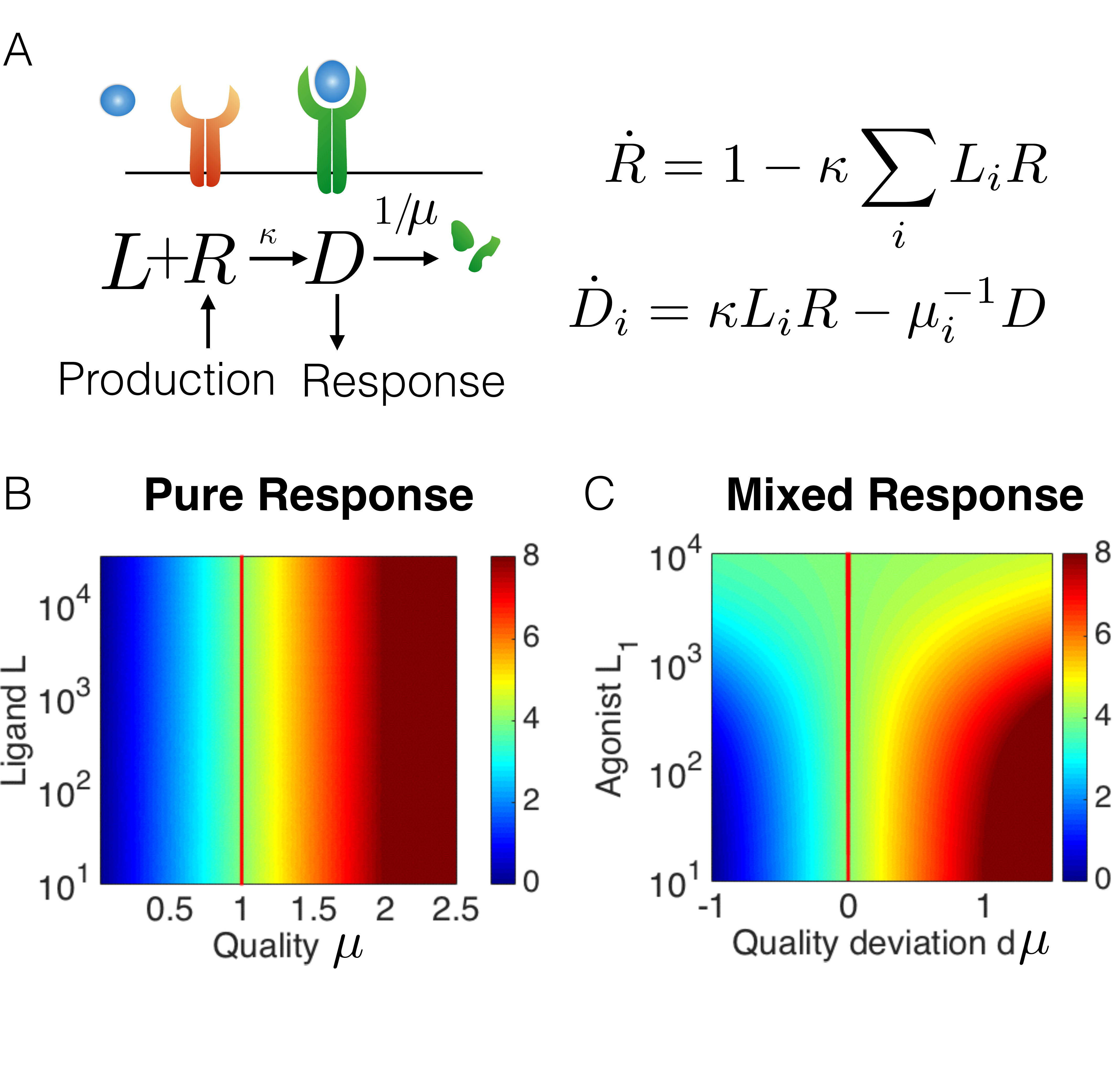}
\caption{(A). Sketch of homeostatic network considered in main text, with the two corresponding equations. There are only three parameters: production rate of $R$ (rescaled to $1$), ligand-receptor association rates $\kappa$ (same for all ligands) and degradation rates of ligand-receptor complex ($\mu^{-1}$, defining ligand quality). For simulations we use  $\kappa=10^{-3}$, and define $\mu_c=\Theta=4$.  (B). Color map of output $T$ values in the $(L,\mu)$ plane,  ligand quality is plotted in units of $\mu_c$. Discrimination line (solid red) is vertical similar to figure \ref{fig:f1} A (C) Color map of output $T$ values in plane $(L_1,d\mu)$ for mixtures of $L_2=10^3$ ligands, quality  $\mu+ d\mu$, with $L_1$ ligands, quality $\mu_c$. This plane is orthogonal to the plane displayed in panel B. Threshold  line is vertical at $d\mu=0$ similar to figure \ref{fig:f1} B. The existence of a region without response in this panel despite high $L_1$ when $d\mu<0$ is indicative of antagonism.}  \label{fig:f1p}
\end{figure}

 A ``homeostatic model'' (Figure \ref{fig:f1p}) is inspired by a ligand-receptor adaptive network evolved in \cite{Francois:2008}, and implements both absolute discrimination and linear antagonism at all orders as described above. Receptors are produced with fixed rate (rescaled to $1$ without loss of generality). Receptor-ligand complexes ($D_{i}$) can form irreversibly with association rates $\kappa$ but are degraded inside the cells with time-scale $\mu_i$, defining quality of ligands (see Figure \ref{fig:f1p} A and equations in Appendix). Output is the total sum of ligand-receptor complexes. Steady state equations for homeostatic model is for mixture $\{(L_n,\mu_n)\}$ 
 \begin{equation}
R=\frac{1}{\kappa \sum_n L_n}   \hspace{0.5 cm}  T=\sum_n D_n=\kappa R \sum_n \mu_n L_n  \label{SoluceAdapt}
 \end{equation}
 
where we defined the output variable $T$ as the sum of all possible ligand-receptor complexes $\sum_n D_n$.
When only one type of ligand $(L,\mu_c)$ is present, $T=\sum_n D_n=\mu_c$ is adaptive (i.e. independent from $L$) as expected. At steady state, $R$ is inversely proportional to total ligand concentration presented irrespective of their $\mu$s. Importantly, this is not due to a titration effect of a fixed pool: $R$ {\it dynamically} buffers any ligand addition, so that steady state output $T$  is a weighted linear combination of $\mu_n$s in \ref{SoluceAdapt}. Such combination is always inferior to the maximum of $\mu_n$, so if such maximum indeed is $\mu_c$,  antagonism ensues.   If we consider the mixture of $L_1$ ligands ( critical time $\mu_c$) with $L_2$ ligands ($\mu_c-d \mu$) we get
 \begin{equation}
 T= \mu_c - d\mu \frac{L_2}{L_1+L_2} \quad \mathrm{i.e. } \hspace{0.5 cm}\mathcal{A}(L,\mu_c)= L^{-1} \label{AntaAdapt}
 \end{equation}
 This contribution is exact and linear in $d\mu$ at all orders. Note that $R$ variable directly  encodes biochemically the antagonistic strength $\mathcal{A}$ (modulo the $\kappa$ prefactor that sets up time-scale but compensates in steady state expression).

Figure \ref{fig:f1p}B and C show values of $T$ for pure ligands and mixture, with vertical discrimination lines at $\mu_c$. This simple model thus represents a perfect implementation of the principles sketched in Figure \ref{fig:f1}. Surprinsingly, absolute discrimination for this model works at steady state even in the limit $L\rightarrow 0$, however it should be noted from equations in  Figure \ref{fig:f1p} that the speed of convergence slows down proportionally to $L$ in this limit, so that for small $L$ one essentially never reaches steady state.

Antagonism strength is proportional to the deviation of antagonists' quality $d\mu$ from $\mu_c$ (equation \ref{AntaAdapt}) . In particular, antagonism is very weak for ligand quality close to $\mu_c$ and gets stronger as quality of ligands gets further below threshold.  Interestingly the hierarchy of antagonism with increasing antagonism for lower quality $\mu$ actually is completely opposite to antagonism observed in immune examples. It is known there that antagonism is maximum for ligands close to critical quality, and vanishes for very small binding time, corresponding to self \cite{AltanBonnet:2005}. It is a biological necessity that self ligands, which are the most frequent ones, do not antagonize immune response (otherwise the immune system would not be able to detect any foreign agent), so it is not clear a priori how to reconcile this with the constraint that absolute discrimination implies antagonism, and a special study in such context is thus required.

\section*{Proofreading based models}

Many immune models are based on kinetic proofreading, first proposed in this context by McKeithan  \cite{Mckeithan:1995}. This model was based on the observation that  one of the main difference between self and not-self ligands is their binding time $\tau$ to the T cell receptors. Kinetic proofreading in this context assumes that a ligand receptor complex can undergo subsequent steps of phosphorylations, but that upon release of the ligand, phosphorylation state of the receptor is reinitialized. It is well known since the original proposal of kinetic proofreading by Hopfield and Ninio \cite{Hopfield:1974,Ninio:1975} that each of those phosphorylation steps can contribute geometrically up to a factor $\tau$, so that the steady state number of the last complexes in a proofreading cascade with $N$ steps in the absence of saturation is roughly proportional to $L\tau^N$, with $L$ the number of ligands  presented. Following the observation of sequential phosphorylations on internal tails of immune receptors \cite{Kersh:1998a}, many current models for immune recognition \cite{Davis:2006,AltanBonnet:2005, Francois:2013,Dushek:2014dv} have elaborated on these first ideas developed by McKeithan, see also \cite{Lever:2014} for a review/comparison between models.

 ``Life-time dogma"  \cite{Feinerman:2008b} posits that immune recognition is an almost absolute discrimination process where the quality of ligand $\mu$ is encoded by binding time of ligands $\tau$ \cite{Voisinne:2015}. This notion first came from a series of experimental data suggesting that the typical ligand concentration to trigger response falls by at least 4 to 5 orders of magnitude with a moderate change of binding time $\tau$ \cite{Kersh:1998wx,Gascoigne:2001}. Various theoretical and experimental studies of this behaviour have led authors to propose that kinetic proofreading should be complemented with combinations of internal positive/negative feedbacks to explain such observed high specificity in $\tau$  \cite{AltanBonnet:2005,Feinerman:2008b,Francois:2016ig}.

\begin{figure}[h]
    \centering
    \includegraphics[width=0.8\textwidth]{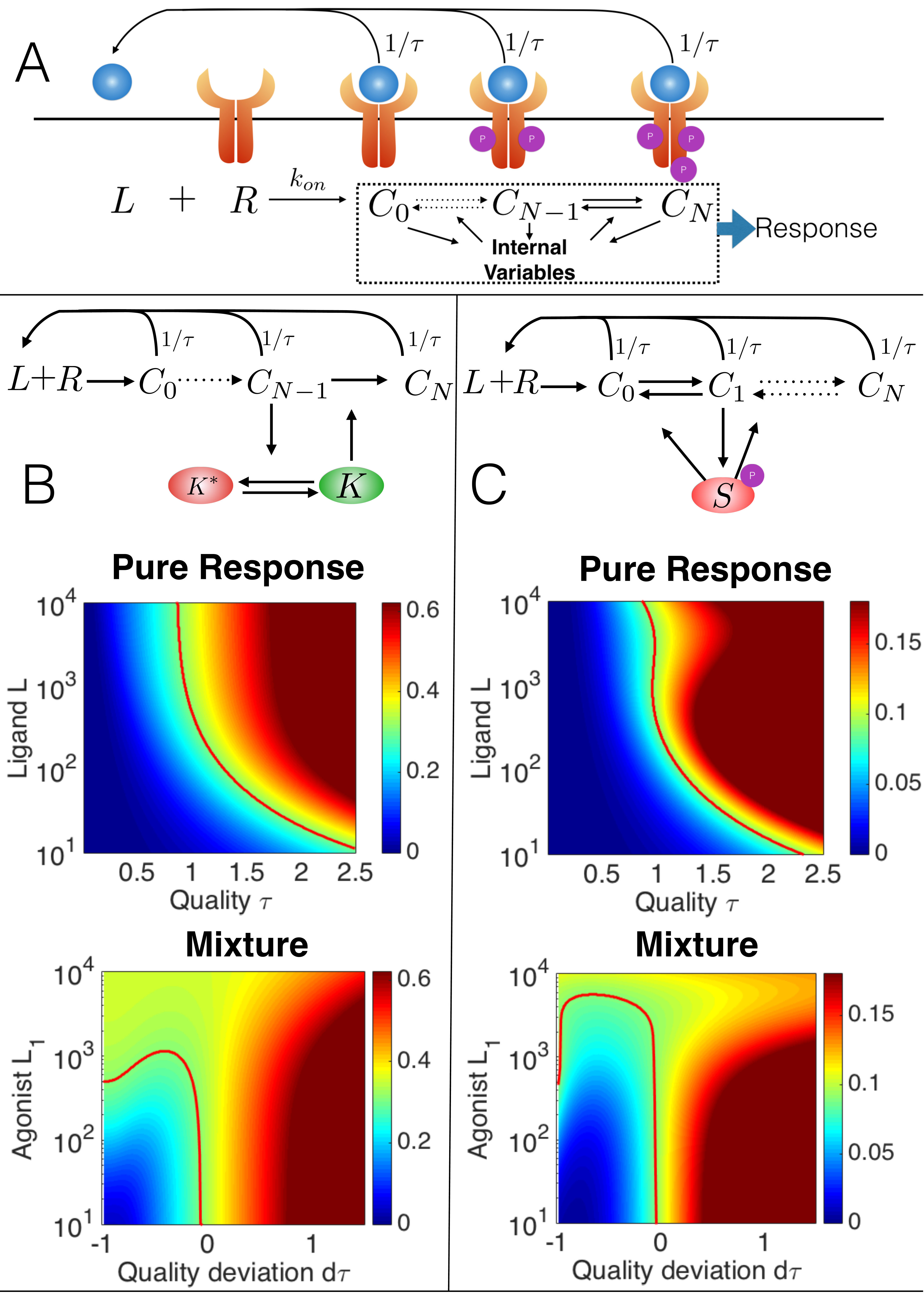}
   \caption{(A) General topology considered for proofreading-based models, with shared internal variables ${\bf M}$. (B) Adaptive sorting topology and response, with colormap for output $C_N$ for pure ligands and mixtures. Here kinase K activity is downregulated by $C_{N-1}$, which also is the substrate of the corresponding phosphorylation .  Equations and parameters are in Appendix, see also the main text for a more detailed description of the model.  We fix $L_2=10^3$ ligands for mixtures, using similar conventions to Figure \ref{fig:f1p}C. Threshold $\Theta$ for discrimination line was chosen so that response is triggered by $L_1 \sim 500$ ligands at $\tau_c$. For  $d\tau \sim -1$, downward folding of both response line and green region for mixture indicates decreased antagonism  (compared to Figure \ref{fig:f1p} C). (C) Immune model  with same conventions as (B), equations and parameters are in Appendix, see main text for description of the model. Phosphatase $S$ is activated by $C_1$ and dephosphorylates all steps. }
    \label{f2}
\end{figure}

 We start with a general derivation for proofreading-based models realizing absolute discrimination, to connect them to biochemical adaptation.   Receptors exist in different   phosphorylation states (notation $C_{j}$s, Figure \ref{f2} A, where $j$ indicates the step in the phosphorylation cascade).  Transition rates between states are assumed to be functions of internal variables (notation $\bf M$), accounting for all signalling inside cells (kinases, phosphatases). Those effectors are assumed to diffuse freely and rapidly, so that each receptor sees the same value for their concentrations.   $\bf M$ values depend on the total occupancy of some $C_j$s. Downstream decision is usually assumed to be effectively controlled by thresholding on one state, index $t$ in the proofreading cascade (i.e. $T=C_t$). Using standard assumptions, it is easy to show (Appendix) that for such proofreading based models, in the limit of unsaturated receptors, the number of receptor states $C_j$ bound to ligands  $(L,\tau)$ takes the functional form:
\begin{equation}
 C_{j}(L,\tau)=c_j({\bf M},\tau) L \label{Separation}
 \end{equation}
  i.e. we can deconvolve influence of ligands into a linear contribution ($L$), and an indirect one ($c_j$) purely due to internal variables.
  
  In the context of absolute discrimination, with quality $\mu=\tau$, equation \ref{Adaptation} imposes that  $T=C_{t}$ is tuned to threshold $\Theta$ irrespective of $L$ at critical quality $\tau_c$. Equation \ref{Separation} thus implies that the indirect contribution $c_t$ to output variable $T$ verifies
        \begin{equation}
            c_{t}({\bf M},\tau_c)=\Theta/L \label{c_t}
       \end{equation}

$c_t$ is a pure function of internal variables, that decreases as a function of $L$. But from equation  \ref{Separation}, $c_t$  contributes multiplicatively to the signal. So this means that  internal variables are used to compensate the ``direct'' linearity of $L$ (from equation \ref{Separation}), thus necessary implementing an incoherent feedforward or feedback loop  \cite{Mangan:2003}  via $c_t$ (which plays here a role similar to $R$ in homeostatic model above).

 This is exactly what happens in the so-called adaptive sorting model (  Figure \ref{f2} B, see full equations in appendix).  In this model, a kinase $K$ responsible for one of the phosphorylation step of the cascade is negatively regulated by an earlier step in the cascade.  Because of this, effective phosphorylation rate mediated by $K$ takes the functional form of equation \ref{c_t}  for high enough  ligands $L$, and as a consequence the concentration of the last complex of the cascade is a pure function of $\tau$ \cite{Lalanne:2013}. Decision can then be made by thresholding on $C_N$ (vertical asymptotic discrimination line on Figure \ref{f2} B).

Another model  with similar behaviour,  inspired by T cell immune recognition network \cite{Francois:2013},  is displayed on  Figure \ref{f2} C (see full equations in appendix). While adaptive sorting relies on repression of an internal kinase $K$, this model instead uses activation of a phosphatase (variable $S$) to provide the incoherent feedback term, allowing for adaptation similar to equation \ref{c_t}. One caveat though is that because of non-specificity of the phosphatase that acts on all steps in the cascade, adaptation is not perfect,  but still antagonism is observed  (Figure \ref{f2} C , see \cite{Francois:2013} for an explicit study)

An important difference with the homeostatic model is that transition rates between variables (and thus $c_j$s) depend explicitly on ligand binding time $\tau$ in proofreading models. It is then informative to consider a slightly more general ansatz with only one internal variable $M$ and separable influence of ligand quality $\mu$,  so that the total output variable  from a mixture of ligands $\{L_n,\mu_n\}$ is

\begin{eqnarray}
T(\{L_n,\mu_n\})&=& \sum_n c(\mu_n,M)L_n= M^{\beta} \sum_n f(\mu_n)   L_n \label{Ansatz}
\end{eqnarray}

with $M=\sum_n g(\mu_n) L_n$ at steady state for mixture $\{L_n,\mu_n\}$.   $\beta=-1$ gives perfect adaptation but different values of $\beta$ can give realistic biological effects such as loss of response at high ligand concentration as observed in \cite{Francois:2013}. We get by direct differentiation:

\begin{equation}
\mathcal{A}(L,\mu_c) =   L^{\beta}\left. \frac{d}{d\mu}\left(fg^\beta\right)\right|_{\mu=\mu_c} \label{Ant_beta}
\end{equation}

This term is very similar to equation \ref{AntaAdapt}, with  an extra $\mu$ dependency coming from $f,g$. If we impose that mixture of agonists trigger response , we necessary have $\mathcal{A}>0$.   If $f\propto \mu^{t+1}$ and $g\propto\mu^{m+1}$ which is approximately the case for adaptive sorting (see Appendix), we thus  have  by substituting in equation \ref{Ant_beta}, $t+1+\beta(m+1)>0$, and thus for $\beta=-1$ we get $t>m$ . In many immune recognition models \cite{AltanBonnet:2005, Francois:2013, Lalanne:2013} this constraint is naturally realized because the internal variable (kinase or phosphatase) implementing approximate biochemical adaptation is regulated by a much earlier step ($m$) than the output ($t$) within the same proofreading cascade. This shows how the structure of proofreading cascades with additional feedback constrains both discrimination and antagonism. Furthermore, as soon as $\beta<0$, in a similar way to $c_t$ above, $\mathcal{A}$ encodes a negative feedforward effect. 

To understand what happens for $\mu<<\mu_c$, it is more illuminating to compute ratio of responses (calling $r$ the fraction of total ligands $L$ with quality $\mu<\mu_c$)
\begin{eqnarray}
\rho&=&\frac{ T(\{(1-r)L,\mu_c;rL,\mu\})}{T(L,\mu_c)}\nonumber\\&=&\left(1-r+\frac{f(\mu)}{f(\mu_c)}r\right)\left(1-r+\frac{g(\mu)}{g(\mu_c)}r\right)^\beta \label{ratio}
\end{eqnarray}

\begin{figure}[h]
    \centering
    \includegraphics[width=0.8\textwidth]{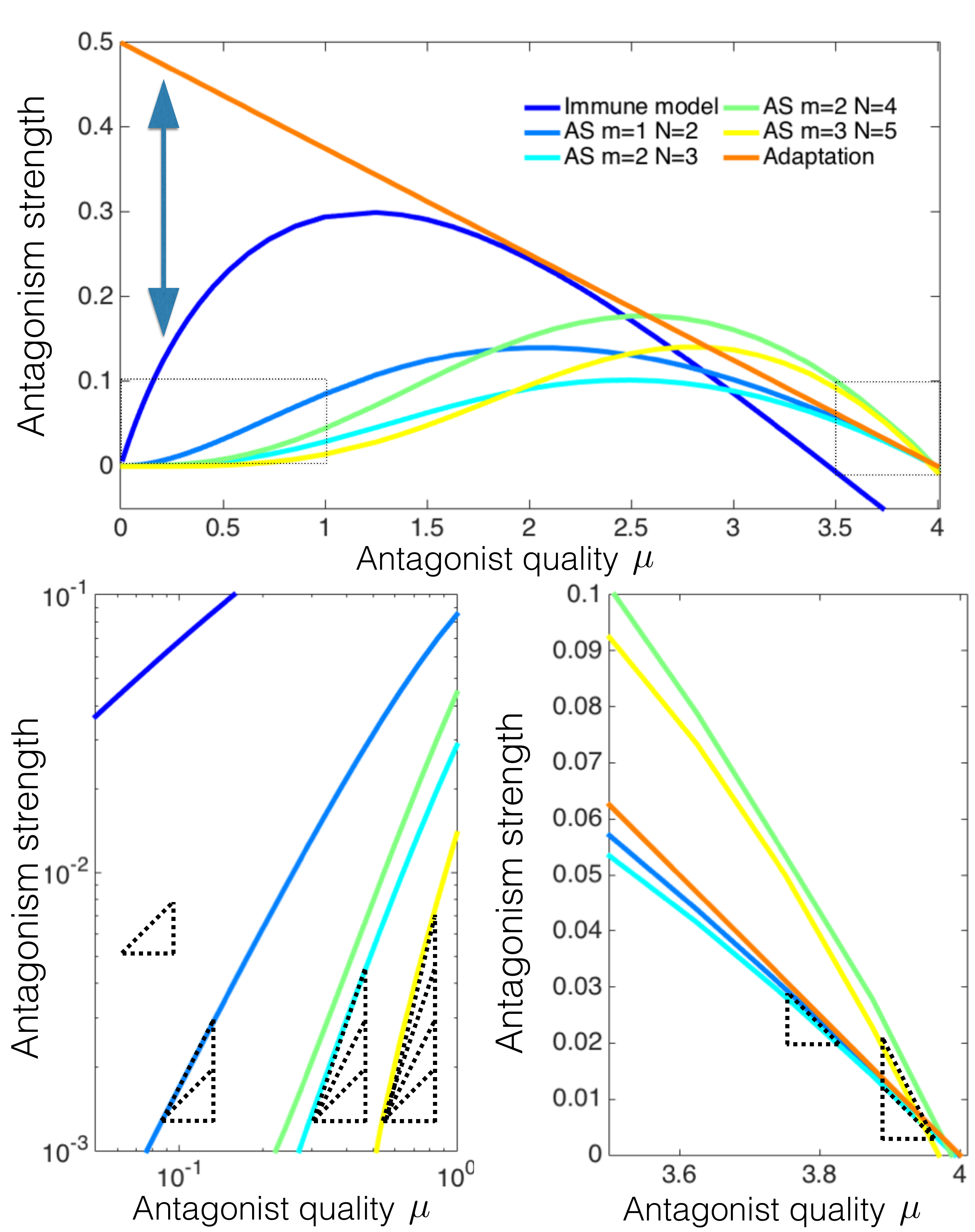}
    \caption{(Top) Computation of relative antagonism strength for different models  with varying parameters. We consider $L_1=L_2=4 \times 10^3$ ligands, $\mu_c=4$, and plot $1-T\{(L_1,\mu_c),(L_2,\mu)\}/T(L_1,\mu_c)$ as a function of $\mu$. For instance, for homeostatic model, antagonism is $r(1- \frac{\mu}{\mu_c})$.  Double arrow indicates region of inverted hierarchy for antagonism. Dotted squared highlight regions plotted on bottom panels. AS indicates adaptive sorting model similar to  Figure \ref{f2} B, Immune model corresponds to model of  Figure \ref{f2} C. For those models, quality $\mu$ is defined by binding time $\tau$. (Bottom left) Log-log plot, showing the dependency of antagonism for small $\mu$. Dashed triangles indicate slopes of $1,2,3,4$. For adaptive sorting models, slope is roughly equal to $m+1$ as expected from the form of $g$. (Bottom right) Linear plot close to $\mu=\mu_c$. Aspect ratio has been chosen so that slope predicted for the homeostatic model $-L_2/(L_1+L_2)\mu_c$ is plotted with slope $-1$. Dashed triangles indicate slopes of $-1,-2$. For adaptive sorting models, slopes are thus roughly equal to $-(N-m)L_2/(L_1+L_2)\mu_c$ as expected from the Ansatz. }
    \label{f3}
\end{figure}

In models of immune detection  based on proofreading such as adaptive sorting, $f$ and $g$  are powers of quality $\mu=\tau$ (binding time), so that $f(\mu)/f(\mu_c),g(\mu)/g(\mu_c) <<1$ when $\mu<\mu_c$. So if $\beta=-1$, we see that $\rho\rightarrow 1$ when $\mu/\mu_c\rightarrow 0$, explaining why antagonism disappears in this parameter regime (corresponding to self in immune detection).  Proofreading models interpolate between  no antagonism for $\mu=\tau\rightarrow 0$ and linear antagonism close to threshold. Thus antagonism strength increases as $\tau$ increases from $0$, as observed experimentally \cite{AltanBonnet:2005}, before quickly collapsing again very close to threshold. Quality of ligands $\tau$ for maximum antagonism is closer to threshold $\tau_c$ with increasing $m$ and $t-m$ (Figure \ref{f3} ).

Interestingly, far from threshold, behaviour of antagonism with respect to quality $\mu$ is a power of $m+1$, i.e. the negative internal contribution $g$ dominates, while close to threshold, the power $t-m$ dominates, which quantifies the competition between the direct positive ($f$) and indirect negative ($g$) contribution to signal. So practically, measurement of antagonism for different $\mu$s provides a way to quantify direct and indirect contributions to response, which suggests an experimental strategy to quantify feedbacks inside such systems via measurements of antagonism.

Finally, if $\beta$ is not exactly equal to $-1$, the system is not perfectly adaptative, but positivity of $\mathcal{A}$ still implies antagonism of similar magnitude to models with perfect adaptation. Assuming a power-law dependency so that $T\propto L^{\epsilon}$ close to $\mu=\mu_c$, then one can show (see Appendix) that upon addition of sub threshold ligands with quality $\mu_c-d\mu$, antagonism appears as soon as $d\mu$ is of order $\epsilon$ (in proper units). This can be seen on the model of \cite{Francois:2013}, illustrated on Fig. \ref{f3} A ``Immune model". For ligand quality close to rescaled quality $1$, there is  a synergestic effect (represented here by negative antagonism) due to the  small increase of response upon addition of ligands of same quality ($\epsilon$ power-law). However, as soon as the quality of the $L_2$  ligands is sufficiently low, we see a decrease in response linear in $d\mu$, until one reaches the antagonistic regime  when the (negative) effect of lowering ligand quality ($\mathcal{A} d\mu$) dominates the (positive) effect of ligand concentration increase ($\epsilon \log (1+\frac{L_2}{L_1})$).

 \section*{Discussion}

In conclusion, we have established how antagonism is a necessary by-product of  strict specific and sensitive detection, thus qualifying as a ``phenotypic spandrel".   While there have been recent studies of spandrels at the level of protein structure \cite{Manhart:2015eg}, to our knowledge, this is the first study of an explicit spandrel structuring a complex signalling phenotype.

With the exception of \cite{Siggia:2013}, our modelling hypotheses correspond to most phenotypic models of immune ligand discrimination we are aware of \cite{Lever:2014}, as well as more complex models such as the one in \cite{AltanBonnet:2005}. Noise can be included as in \cite{Francois:2013,Lalanne:2013}  via time-integration of some output variable, replacing all concentrations by expectation values, e.g. $C \rightarrow <C>$ .  To possibly disentangle antagonism from underlying biochemical adaptation one would need to examine limiting cases or complexify hypotheses.  One could consider for instance the time-course of a variable far from steady-state   \cite{Siggia:2013} or coupled to multi stability \cite{Lipniacki:2008} to ``break'' the differentiability hypothesis.

Antagonism appears as direct consequence of biochemical adaptation. Further relaxing assumptions to look for alternative mechanisms, it is useful to write Taylor expansion of  output $T(L,\mu)$ for pure ligands close to $\mu=\mu_c$:

\begin{equation}
dT=\frac{\partial T}{\partial L} dL+\frac{\partial T}{\partial \mu} d\mu \label{dT}
\end{equation}

As said before, strict specificity imposes that for at $\mu=\mu_c$, $T=\Theta$ independently of $L$, which mathematically implies from equation \ref{dT} that for $d\mu=0$, $dT=0$ so that $\left. \frac{\partial T}{\partial L}\right|_{\mu=\mu_c}=0$, which is biochemical adaptation as assumed throughout this manuscript. Imperfect adaptation means non zero $ \frac{\partial T}{\partial L}$, but if it is small enough we saw previously that we still expect antagonism. Notice in that case from equation \ref{dT}  that $dT$ a priori stays small as both $dL$ and $d\mu$ vary. But when we relax the strict specificity hypothesis so that $dT=0$ for small but non zero $d\mu$, another possible limit appears, most visible by computing  the slope of the discrimination line (for $dT=0$) :

\begin{equation}
\frac{d L}{d\mu}=-\frac{\frac{\partial T}{\partial \mu}}{\frac{\partial T}{\partial L}}
\end{equation}

To get a vertical discrimination line in the $(L,\tau)$ plane, this slopes need to go to $\infty$, so one could either take $\frac{\partial T}{\partial L}=0$ in the denominator (adaptation), or directly take $\frac{\partial T}{\partial \mu} \rightarrow \infty$ corresponding to infinite kinetic amplification. This could be approximated by kinetic proofreading with high number $N$ of proofreading steps, where an amplification up to $\tau^N$ in magnitude can be obtained.  For instance, in a model explicitly designed to provide better sensitivity \cite{Dushek:2014dv}, a rather high number of steps ($N=25$) is taken to account for specificity compatible with biological ranges, and indeed such models do not yield antagonism.  Notice however that if $\frac{\partial T}{\partial \mu} \rightarrow \infty$, $dT$ in equation \ref{dT} varies very rapidly when $d\mu$ changes. So in this limiting case,  the output itself is (by construction) infinitely sensitive to changes of $\mu$. This is why there is no antagonism: subthresholds ligands yield infinitely small contributions compared to the ones at threshold, irrespective of their concentration, and in particular can not antagonize them.  As pointed out in \cite{AltanBonnet:2005}, depending on the system considered, it is actually not clear that many proofreading steps are available to the cell. This observation led to the proposal that specificity is rather due to the presence of internal feedbacks, effectively reducing instead $\frac{\partial T}{\partial L}$ to perform specific detection, which is the framework of this article.

At least two examples of absolute discrimination/antagonism are offered in the immune context. For early immune recognition mediated by TCRs (discrimination parameter being binding time $\tau$), agonists simultaneously presented with ligands just below threshold fail to trigger response, while agonists alone do  \cite{Dittel:1999,AltanBonnet:2005}. There is some debate in immunology about the parameter defining ligand quality: some argue that the binding time $\tau$ defines quality, while some other authors \cite{Dushek:2009fi,Govern:2010kx,Aleksic:2010hh} defined an effective parameter (e.g. dwelling time $t_a$ in  \cite{Govern:2010kx}) to encode ligand strength.  But if all ligands can be hierarchically ordered on one ``quality" axis defining  effectively $\mu$, whether $\mu=\tau$ or $\mu=t_a$,  our reasoning applies and antagonism should ensue.   For  Fc$\epsilon$RIs mediated response in mast cells \cite{Torigoe:1998vj}, where discrimination parameter also is binding time $\tau$, a very similar ``dog in the manger" effect is observed where low affinity antagonists titrate the Lyn kinase responsible for proofreading steps (exactly like adaptive sorting evolved in \cite{Lalanne:2013}).  

Possible  instances outside of immunology include antagonism via negative feedback in Hh signalling \cite{Holtz:2013dj}  or hormonal pathways, which present properties very reminiscent of immune recognition \cite{Francois:2013}.  Non-monotonic response activity with varying ligand concentration \cite{Vandenberg:2012fc,Fagin:2012dw}, could correspond in our framework to approximate adaptation where $\frac{\partial T}{\partial L}$ is kept small. It has been established that in such context antagonists differ from agonists purely based on slower binding kinetics \cite{Rich:2002ju},  suggesting a hierarchical ordering of ligand quality as hypothesized here. A recent meta-study shows that most of these pathways  indeed use internal negative feedbacks to change monotonicity of response \cite{Lagarde:2015cw}, thus similar to the negative feedforward contributions predicted from equations \ref{c_t} and \ref{Ant_beta}.

Is antagonism an evolutionary spandrel related to immune detection ? Networks  evolved in \cite{Lalanne:2013} systematically show local biochemical adaptation and antagonism, as expected from the present derivation. So evolutionary spandrels can be observed and studied in evolutionary simulations, and their emergence studied theoretically. In nature, it is  difficult to definitely know if a trait is a spandrel without detailed historical data and comparisons \cite{Gould:1997ue}. Other possibilities could be that absolute discrimination has been selected in conjunction with other properties (such as information transmission \cite{Mora:2015cv,Singh:2015vs} or statistical decision \cite{Lalanne:2015}). But as detailed mathematically in this work,  the presence of phenotypic spandrel here is due to the non-trivial computation performed by the cell to disentangle  kinetic of binding and ligand concentration. Experiments probing internal feedbacks (such as  \cite{Torigoe:1998vj}) connect spandrels to cellular computations, similar to symmetries connected to physical laws.

{\bf Acknowledgements}. We thank Jean-Beno\^it Lalanne and Gr\'egoire Altan-Bonnet for useful discussions. This project has been funded by the Natural Science and Engineering Research Council of Canada (NSERC), Discovery Grant Program, and a Simons Investigator Award for Mathematical Modelling of Biological Systems to PF. LS has been supported by an Undergraduate Summer Research Award from NSERC. 

\section*{Appendix}

\subsection*{Equations for adaptive sorting (AS)}

\begin{eqnarray}
K=&K_T/(C_m+C^*)   \label{K} \\
\dot C_0=&-(\phi_0+\tau^{-1} ) C_0+\kappa (L-\sum_i C_i) (R- \sum_i C_i)  \\ 
\dot C_i=&-(\phi_i+\tau^{-1} ) C_i+\phi_{i-1} C_{i-1}, \hspace{0.15cm} 1\leq i    
\end{eqnarray}

 where  $\phi_m=\phi_K K$, $\phi_N=0$ and $\phi_i=\phi$ for other values of $i$.

Parameters used for simulation in main text are $\kappa=10^{-5}, R=3 \times 10^4, \phi=\phi_K K_T=0.09, C^*=3, m=2, N=3$. Output is $C_N$, i.e. with conventions defined in the main text $t=N$. We used $\tau_c= 4 \ s$ and defined threshold $\Theta=0.31$ to plot a discrimination line.

We get immediately at steady state

\begin{equation}
C_j=\tau^j  \lambda_j C_0
\end{equation}

setting $\gamma_j=\frac{\phi_{j-1}}{\phi_{j} \tau+1}$ and  $\lambda_0=1,\lambda_j=\Pi_{1\leq k \leq j} \gamma_k$. Neglecting $\sum_i C_i $ in front of $R$ we have

\begin{equation}
C_0=\frac{\kappa R L \tau}{ \kappa R \tau \sum_j \lambda_j \tau^j + \phi_0 \tau +1   } \label{C0}
\end{equation}

Assuming  $\phi_j \tau, \kappa R \tau<<1$, we then get the following scaling laws for $i \leq m$

\begin{equation}
C_i=\kappa R \tau^{i+1} \phi^{i} L 
\end{equation}

and for $i>m$

\begin{equation}
C_i=\kappa R \tau^{i+1}  \phi^{i-1} \phi_K K  L
\end{equation}

Since $K\propto C_m^{-1}$, we recover scaling  law of the Ansatz from the main text.

\subsection*{Equations for immune model}

\begin{eqnarray}
\label{eq1}
S&=&S_T \frac{C_1}{C_1+C_*}\\
\label{eq2}
\dot{C_0}&=&\kappa (L-\sum_{i=0}^N C_i)(R-\sum_{i=0}^N C_i) \nonumber \\&&+(b+\gamma S)C_1-(\phi+\tau^{-1} )C_0\\  \nonumber
\label{eq3}
\dot C_j&=&\phi C_{j-1}+(b+\gamma S)C_{j+1}-(\phi+b+\gamma S+\tau^{-1} )C_j\\  \nonumber
\label{eq4}
\dot C_N&=&\phi C_{N-1}-(b+\gamma S+\tau^{-1} )C_N
\end{eqnarray}

Parameters used for simulation in main text are $\kappa=10^{-4}, R=3 \times 10^4, \phi=0.09, b=0.04, \gamma S_T=0.72, C_*=300$.  Output is $C_N$. We used $\tau_c= 4 \ s$ and defined threshold $\Theta=0.09$ to plot a  discrimination line.

\subsection*{Origin of linear separation (equation \ref{Separation})}

Considering a  family of models similar to the ones defined above (e.g. the ones in \cite{Lever:2014}), it is clear that if we assume unsaturated receptors, i.e. $\sum_{i=0}^N C_i<<R$, calling $\bf C$ the vector of occupancies and ${\bf M}$ the internal variables , we have:

\begin{equation}
\dot {\bf C}=\kappa({\bf M}) R  {\bf L}+ \mathcal{T}(R,{\bf M})  {\bf C} \label{system}
\end{equation}

${\bf L}=(L,0,\dots,0)$ is the vector corresponding to ligand input, $\kappa ({\bf M})$ the association rate of ligands to receptors - by definition here ligands and receptors bind into the first state $C_0$. $\mathcal{T}(R,{\bf M})$ is a matrix defining linear rates between occupancies states, depending on internal variables and parameter $\tau$. Dynamics and steady state value of ${\bf M}$ is given by extra equations that are model-specific (e.g. equations \ref{K} and \ref{eq1} above).

For such systems, we have at steady state

\begin{equation}
{\bf C}= - \kappa({\bf M}) R \mathcal{T}(R,{\bf M}) ^{-1}{\bf L}
\end{equation}
from which we can directly compute the $c_j$s  as a function of ${\bf M}$ to get a functional form similar to equation \ref{Separation}. Then we can use equations defining steady-state values of ${\bf M}$ as function of {\bf C} to close the system.

When several ligand types are present, independence of ligand binding means that a system similar to equation \ref{system} holds for every single vector of occupancy ${\bf C}^\tau=(C^\tau_j)$ of receptor states bound to ligands with quality $\tau$. Coupling between different types of ligands only occur via internal variables {\bf M}. Note that we can also generalize this formalism so that transition rates depend on occupancies (i.e. effectively giving non linear transition rates between states) by assuming that internal variables ${\bf M}$ are occupancies themselves. The underlying strong assumption here is that the coupling is global, via total occupancies.

\subsection*{Antagonism when adaptation is not perfect}

In this section, we briefly illustrate what happens to antagonism when adaptation is not perfect, using the Ansatz presented in the main text as a case-study. As said in the main text, antagonism then relies on a competition between the ``flatness" of $T$ as a function of $L$ and variation of $d\mu$.
To see this, let us consider the Ansatz of the main text with $\beta=-1+\epsilon$, with $\epsilon>0$ so that  $T(L,\mu_c) \propto L^\epsilon$ slowly increases as a function of $L$ for $\mu=\mu_c$. Writing the equivalent of equation  \ref{Taylor_anta}  to get:

\begin{eqnarray}
T(\{L_1,\mu_c;L_2,\mu_c-d\mu\})&=& (L_1+L_2)^{\epsilon}\mathcal{B}(\mu_c) \\ \nonumber
&-&L_2 (L_1+L_2)^{-1+\epsilon} d\mu \mathcal{B}'(\mu_c) 
\end{eqnarray}

calling   $\mathcal{B}=fg^\beta$. Then for the difference of output $\Delta T=  T(\{L_1,\mu_c;L_2,\mu_c-d\mu\})-T(L_1,\mu_c)$ we get

  \begin{eqnarray}
\Delta T&=&(  (L_1+L_2)^{\epsilon}-L_1^{\epsilon})\mathcal{B}(\mu_c)\\ \nonumber
  &-& L_2 (L_1+L_2)^{-1+\epsilon}d\mu \mathcal{B}'(\mu_c) \\
  &\simeq & \epsilon \log (1+\frac{L_2}{L_1})\mathcal{B}(\mu_c)- L_2 (L_1+L_2)^{-1+\epsilon}d\mu \mathcal{B}'(\mu_c) \label{anta_approx}
  \end{eqnarray}
  
One can see that the first term in equation \ref{anta_approx} is of order $\epsilon$, while the second term is the usual negative antagonistic  term of order $d\mu$, and in the limit of small $L_2$ compared to $L_1$, both terms are linear in $L_2$.  So for $d\mu$ big enough compared to $\epsilon$, the second term should dominate the first one, and we expect antagonism to occur even without perfect adaptation. This can be observed on an even less general model with non-monotonic dose response curve from \cite{Francois:2013}, illustrated on Figure \ref{f2} C, where the output can vary over one decade while the input varies over 4 decades  \cite{Francois:2013}, but still the system displays antagonism as soon as $\tau$ is sufficiently below  $\tau_c$.

\bibliographystyle{unsrt}

\begin{thebibliography}{10}

\bibitem{Shoval:2012ke}
O~Shoval, H~Sheftel, G~Shinar, Y~Hart, O~Ramote, A~Mayo, E~Dekel, K~Kavanagh,
  and Uri Alon.
\newblock {Evolutionary trade-offs, Pareto optimality, and the geometry of
  phenotype space.}
\newblock {\em Science}, 336(6085):1157--1160, June 2012.

\bibitem{Warmflash:2012}
Aryeh Warmflash, Paul Fran{\c c}ois, and Eric~D Siggia.
\newblock {Pareto evolution of gene networks: an algorithm to optimize multiple
  fitness objectives.}
\newblock {\em Physical Biology}, 9(5):056001--056001, October 2012.

\bibitem{Francois:2014}
Paul Fran{\c c}ois.
\newblock {Evolving phenotypic networks in silico.}
\newblock {\em Seminars in cell {\&} developmental biology}, 35:90--97,
  November 2014.

\bibitem{Gould:1979vo}
S~J Gould and R~C Lewontin.
\newblock {The spandrels of San Marco and the Panglossian paradigm: a critique
  of the adaptationist programme.}
\newblock {\em Proceedings of the Royal Society of London. Series B, Containing
  Papers of a Biological Character}, 205(1161):581--598, September 1979.

\bibitem{Gould:1982ta}
S~J Gould and E~S Vrba.
\newblock {Exaptation-a missing term in the science of form}.
\newblock {\em Paleobiology}, 1982.

\bibitem{Gould:1997ue}
S~J Gould.
\newblock {The exaptive excellence of spandrels as a term and prototype.}
\newblock {\em Proceedings of the National Academy of Sciences of the United
  States of America}, 94(20):10750--10755, September 1997.

\bibitem{Malissen:1998dy}
B~Malissen.
\newblock {Translating Affinity into Response}.
\newblock {\em Science}, 281(5376):528--529, July 1998.

\bibitem{Feinerman:2008b}
Ofer Feinerman, Ronald~N Germain, and Gr{\'e}goire Altan-Bonnet.
\newblock {Quantitative challenges in understanding ligand discrimination by
  $\alpha$$\beta$ T cells}.
\newblock {\em Molecular Immunology}, 45(3):619--631, February 2008.

\bibitem{Lever:2014}
Melissa Lever, Philip~K Maini, P~Anton van~der Merwe, and Omer Dushek.
\newblock {Phenotypic models of T cell activation.}
\newblock {\em Nature Reviews Immunology}, 14(9):619--629, September 2014.

\bibitem{Chakraborty:2014hw}
Arup~K Chakraborty and Arthur Weiss.
\newblock {Insights into the initiation of TCR signaling}.
\newblock {\em Nature immunology}, 15(9):798--807, August 2014.

\bibitem{Kersh:1998a}
E~N Kersh, A~S Shaw, and Paul~M Allen.
\newblock {Fidelity of T cell activation through multistep T cell receptor zeta
  phosphorylation.}
\newblock {\em Science}, 281(5376):572--575, July 1998.

\bibitem{Gascoigne:2001}
N~R Gascoigne, T~Zal, and S~M Alam.
\newblock {T-cell receptor binding kinetics in T-cell development and
  activation.}
\newblock {\em Expert reviews in molecular medicine}, 2001(06):1--17, February
  2001.

\bibitem{Dushek:2009fi}
Omer Dushek, Raibatak Das, and Daniel Coombs.
\newblock {A role for rebinding in rapid and reliable T cell responses to
  antigen.}
\newblock {\em PLoS Comput Biol}, 5(11):e1000578, November 2009.

\bibitem{Govern:2010kx}
Christopher~C Govern, Michelle~K Paczosa, Arup~K Chakraborty, and Eric~S
  Huseby.
\newblock {Fast on-rates allow short dwell time ligands to activate T cells.}
\newblock {\em Proc Natl Acad Sci U S A}, 107(19):8724--8729, May 2010.

\bibitem{Aleksic:2010hh}
Milos Aleksic, Omer Dushek, Hao Zhang, Eugene Shenderov, Ji-Li Chen, Vincenzo
  Cerundolo, Daniel Coombs, and P~Anton van~der Merwe.
\newblock {Dependence of T Cell Antigen Recognition on T Cell Receptor-Peptide
  MHC Confinement Time}.
\newblock {\em Immunity}, 32(2):163--174, February 2010.

\bibitem{Francois:2016ig}
Paul Fran{\c c}ois and Gr{\'e}goire Altan-Bonnet.
\newblock {The Case for Absolute Ligand Discrimination: Modeling Information
  Processing and Decision by Immune T Cells}.
\newblock {\em Journal of Statistical Physics}, 162(5):1130--1152, 2016.

\bibitem{Voisinne:2015}
Guillaume Voisinne, G~B Nixon, Anna Melbinger, G~Gasteiger, Massimo Vergassola,
  and Gr{\'e}goire Altan-Bonnet.
\newblock {T Cells Integrate Local and Global Cues to Discriminate between
  Structurally Similar Antigens}.
\newblock {\em Cell reports}, 11(5):1--12, May 2015.

\bibitem{Lalanne:2013}
Jean-Beno{\^\i}t Lalanne and Paul Fran{\c c}ois.
\newblock {Principles of adaptive sorting revealed by in silico evolution.}
\newblock {\em Physical Review Letters}, 110(21):218102, May 2013.

\bibitem{Torigoe:1998vj}
C~Torigoe, J~K Inman, and H~Metzger.
\newblock {An unusual mechanism for ligand antagonism.}
\newblock {\em Science}, 281(5376):568--572, July 1998.

\bibitem{AltanBonnet:2005}
Gr{\'e}goire Altan-Bonnet and Ronald~N Germain.
\newblock {Modeling T cell antigen discrimination based on feedback control of
  digital ERK responses.}
\newblock {\em PLoS Biology}, 3(11):e356, November 2005.

\bibitem{Dittel:1999}
B~N Dittel, I~Stefanova, R~N Germain, and C~A Janeway.
\newblock {Cross-antagonism of a T cell clone expressing two distinct T cell
  receptors}.
\newblock {\em Immunity}, 11(3):289--298, September 1999.

\bibitem{Francois:2013}
Paul Fran{\c c}ois, Guillaume Voisinne, Eric~D Siggia, Gr{\'e}goire
  Altan-Bonnet, and Massimo Vergassola.
\newblock {Phenotypic model for early T-cell activation displaying sensitivity,
  specificity, and antagonism.}
\newblock {\em Proc Natl Acad Sci U S A}, 110(10):E888--97, March 2013.

\bibitem{Mckeithan:1995}
T~W MCKEITHAN.
\newblock {Kinetic proofreading in T-cell receptor signal transduction.}
\newblock {\em Proceedings of the National Academy of Sciences of the United
  States of America}, 92(11):5042--5046, May 1995.

\bibitem{Francois:2008}
Paul Fran{\c c}ois and Eric~D Siggia.
\newblock {A case study of evolutionary computation of biochemical adaptation.}
\newblock {\em Physical Biology}, 5(2):26009, 2008.

\bibitem{Ma:2009}
Wenzhe Ma, Ala Trusina, Hana El-Samad, Wendell~A Lim, and Chao Tang.
\newblock {Defining network topologies that can achieve biochemical
  adaptation.}
\newblock {\em Cell}, 138(4):760--773, August 2009.

\bibitem{Vandenberg:2012fc}
L~N Vandenberg, T~Colborn, T~B Hayes, J~J Heindel, D~R Jacobs, D~H Lee,
  T~Shioda, A~M Soto, F~S vom Saal, W~V Welshons, R~T Zoeller, and J~P Myers.
\newblock {Hormones and Endocrine-Disrupting Chemicals: Low-Dose Effects and
  Nonmonotonic Dose Responses}.
\newblock {\em Endocrine Reviews}, 33(3):378--455, June 2012.

\bibitem{Hopfield:1974}
J~J Hopfield.
\newblock {Kinetic proofreading: a new mechanism for reducing errors in
  biosynthetic processes requiring high specificity}.
\newblock {\em Proceedings of the National Academy of Sciences of the United
  States of America}, 71(10):4135--4139, 1974.

\bibitem{Ninio:1975}
J~Ninio.
\newblock {Kinetic Amplification of Enzyme Discrimination}.
\newblock {\em Biochimie}, 57(5):587--595, 1975.

\bibitem{Davis:2006}
Simon~J Davis and P~Anton van~der Merwe.
\newblock {The kinetic-segregation model: TCR triggering and beyond.}
\newblock {\em Nature immunology}, 7(8):803--809, August 2006.

\bibitem{Dushek:2014dv}
Omer Dushek and P~Anton van~der Merwe.
\newblock {An induced rebinding model of antigen discrimination}.
\newblock {\em Trends in Immunology}, 35(4):153--158, April 2014.

\bibitem{Kersh:1998wx}
Gilbert~J Kersh, Ellen~N Kersh, Daved~H Fremont, and Paul~M Allen.
\newblock {High- and Low-Potency Ligands with Similar Affinities for the TCR}.
\newblock {\em Immunity}, 9(6):817--826, December 1998.

\bibitem{Mangan:2003}
S~Mangan and Uri Alon.
\newblock {Structure and function of the feed-forward loop network motif}.
\newblock {\em Proceedings of the National Academy of Sciences of the United
  States of America}, 100(21):11980--11985, October 2003.

\bibitem{Manhart:2015eg}
Michael Manhart and Alexandre~V Morozov.
\newblock {Protein folding and binding can emerge as evolutionary spandrels
  through structural coupling}.
\newblock {\em Proc Natl Acad Sci U S A}, 112(6):1797--1802, February 2015.

\bibitem{Siggia:2013}
Eric~D Siggia and Massimo Vergassola.
\newblock {Decisions on the fly in cellular sensory systems.}
\newblock {\em Proc Natl Acad Sci U S A}, 110(39):E3704--12, September 2013.

\bibitem{Lipniacki:2008}
Tomasz Lipniacki, Beata Hat, James~R Faeder, and William~S Hlavacek.
\newblock {Stochastic effects and bistability in T cell receptor signaling}.
\newblock {\em Journal of Theoretical Biology}, 254(1):110--122, September
  2008.

\bibitem{Holtz:2013dj}
A~M Holtz, K~A Peterson, Y~Nishi, S~Morin, J~Y Song, F~Charron, A~P McMahon,
  and B~L Allen.
\newblock {Essential role for ligand-dependent feedback antagonism of
  vertebrate hedgehog signaling by PTCH1, PTCH2 and HHIP1 during neural
  patterning}.
\newblock {\em Development (Cambridge, England)}, 140(16):3423--3434, July
  2013.

\bibitem{Fagin:2012dw}
Dan Fagin.
\newblock {Toxicology: The learning curve.}
\newblock {\em Nature}, 490(7421):462--465, October 2012.

\bibitem{Rich:2002ju}
Rebecca~L Rich, Lise~R Hoth, Kieran~F Geoghegan, Thomas~A Brown, Peter~K
  LeMotte, Samuel~P Simons, Preston Hensley, and David~G Myszka.
\newblock {Kinetic analysis of estrogen receptor/ligand interactions.}
\newblock {\em Proceedings of the National Academy of Sciences of the United
  States of America}, 99(13):8562--8567, June 2002.

\bibitem{Lagarde:2015cw}
Fabien Lagarde, Claire Beausoleil, Scott~M Belcher, Luc~P Belzunces, Claude
  Emond, Michel Guerbet, and Christophe Rousselle.
\newblock {Non-monotonic dose-response relationships and endocrine disruptors:
  a qualitative method of assessment}.
\newblock {\em Environmental Health}, 14(1):13--a106, 2015.

\bibitem{Mora:2015cv}
Thierry Mora.
\newblock {Physical Limit to Concentration Sensing Amid Spurious Ligands}.
\newblock {\em Physical Review Letters}, 115(3):038102, July 2015.

\bibitem{Singh:2015vs}
Vijay Singh and Ilya Nemenman.
\newblock {Accurate sensing of multiple ligands with a single receptor}.
\newblock {\em arXiv.org}, May 2015.

\bibitem{Lalanne:2015}
Jean-Beno{\^\i}t Lalanne and Paul Fran{\c c}ois.
\newblock {Chemodetection in fluctuating environments: Receptor coupling,
  buffering, and antagonism.}
\newblock {\em Proc Natl Acad Sci U S A}, 112(6):1898--1903, January 2015.

\end{thebibliography}

\end{document}